\documentclass[twocolumn,nofootinbib,superscriptaddress]{revtex4-1}
\usepackage{latexsym,epsfig,amssymb, amsmath,nicefrac}

\usepackage{color}

\usepackage[makeroom]{cancel}
  \usepackage{latexsym}
  \usepackage{epsf}
  \usepackage{amssymb}
  \usepackage{graphicx}
  \usepackage{amsmath}
  \usepackage{amsmath,amssymb,amsthm}
  \usepackage{verbatim}
  \usepackage{hyperref}
\renewcommand{\d}{\textrm{d}}

\newcommand{\g}{{\bf g}}
\newcommand{\V}{{\mathcal V}}

\newcommand{\Z}{{\mathbb Z}}

\def\bi{\begin{itemize}}
\def\ei{\end{itemize}}
\def\be{\begin{equation}}
\def\ee{\end{equation}}
\newcommand{\bea}{\begin{eqnarray}}
\newcommand{\eea}{\end{eqnarray}}

\def\lsim{\mathrel{\mathop
  {\hbox{\lower0.5ex\hbox{$\sim$}\kern-0.8em\lower-0.7ex\hbox{$<$}}}}}
\def\gsim{\mathrel{\mathop
  {\hbox{\lower0.5ex\hbox{$\sim$}\kern-0.8em\lower-0.7ex\hbox{$>$}}}}}

\begin{document}

\vspace{1cm}
\title{Warping the Weak Gravity Conjecture}

\author{Karta Kooner}
\email{pykooner@swan.ac.uk}
\affiliation{Department of Physics, Swansea University, Singleton Park, Swansea, SA2 8PP, UK }
\author{Susha Parameswaran}
\email{Susha.Parameswaran@liverpool.ac.uk}
\affiliation{Department of Mathematical Sciences, University of Liverpool, Mathematical Sciences Building
Liverpool, L69 7ZL, UK }
\author{ Ivonne Zavala}
\email{e.i.zavalacarrasco@swansea.ac.uk}
\affiliation{Department of Physics, Swansea University, Singleton Park, Swansea, SA2 8PP, UK }

\begin{abstract}
The Weak Gravity Conjecture, if valid, rules out simple models of Natural Inflation by restricting their axion decay constant to be sub-Planckian.  We revisit stringy attempts to realise Natural Inflation, with a single open string axionic inflaton from D-branes in a warped throat.  We show that warping allows the requisite super-Planckian axion decay constant to be achieved consistently with the Weak Gravity Conjecture.  
However, there is a tension between large axion decay constant and high string scale, where the requisite high string scale is difficult to achieve in \emph{all} attempts to realise large field inflation using perturbative string theory.  We comment on the Generalized Weak Gravity Conjecture in the light of our results.
\end{abstract}	

\maketitle

\section{Introduction}

Cosmological inflation stands strong as the leading mechanism to provide  the seeds that gave rise to the large structure we observe today in the Universe. Precision observations in the Cosmic Microwave Background provide a window into this very early history of the Universe.  The latest results from Planck \cite{Planck15Infla} are in perfect agreement with the simplest inflationary models, driven by the dynamics of a single scalar field rolling down a very flat potential.  Current bounds from Planck/BICEP2 on the scalar to tensor ratio in the CMB power spectrum are $r \lesssim 0.12$ ($95\%$ CL).  Any future detection of tensor modes would have the remarkable implications, via the Lyth relation \cite{Lyth,BL,BRSZ}, that inflation occurred at scales close to the Planck scale:
\be
V_{inf}^{1/4} \approx \left(\frac{r}{0.1}\right)^{1/4} \times 1.8 \times 10^{16} \,\textrm{GeV} \label{E:Vinf}
\ee
and that the inflaton field had super-Planckian excursions:
\be
\frac{\Delta\phi}{M_{Pl}} \gsim 0.25 \times \left(\frac{r}{0.01}\right)^{1/2}\,. \label{E:Deltaphi}
\ee

``Large field'' inflationary models are intriguing not only due to their robust prediction 
of high scale inflation with observable primordial gravitational waves.  They depend sensitively on the degrees of freedom comprising the ultraviolet completion of gravity.  
In particular, Planck suppressed corrections to the slow-roll inflaton potential typically become large when the inflaton varies over super-Planckian scales.  One idea to protect the slow-roll inflaton potential from dangerous quantum corrections is to invoke a shift symmetry in the inflaton field, for instance by identifying the inflaton with a Goldstone boson,  the axion.  The classical example in this vein is Natural Inflation \cite{NI}, now tightly constrained by the latest CMB observations\footnote{Although massive modes during inflation can change the classic NI  predictions by generating a smaller than unity speed of sound bringing the model back to the allowed parameter region, as shown in \cite{AAW}.}. 
In Natural Inflation, the axion enjoys a continuous shift symmetry within the perturbative approximation.  This is broken to a discrete symmetry by non-perturbative effects, which generate a potential of the form:
\be
V(\phi) = V_0 \, \left(1 \pm \cos\left(\frac{\phi}{f}\right)\right)\,, \label{E:NI-V}
\ee
where $f$ is the axion decay constant.  The potential is sufficiently flat for slow-roll inflation provided that $f\gg M_{Pl}$, and as a consequence the axion can undergo super-Planckian field excursions.  The current bound on the Natural Inflation potential (\ref{E:NI-V}), given by PLANCK from the spectral index, $n_s$, is $f/M_{Pl} \gsim 6.8$ ($95\%$ CL)  \cite{Planck15Infla}.
 
To understand whether or not such ideas are viable requires the embedding of large field models in a theory of quantum gravity.  In fact, much interest has recently been generated by the possibility that general features of quantum gravity can constrain inflationary models with observable consequences.  The Weak Gravity Conjecture \cite{WGC} roughly proposes that gravity must be the ``weakest force'' in a quantum theory of gravity, in order to avoid stable black hole remnants.  For example, for a four dimensional theory describing gravity and a $U(1)$ gauge sector with gauge coupling, $\g$, there must exist a state with mass, $m$, which satisfies $m \lesssim M_{Pl} \,\g$.  Moreover, the effective field theory has a new UV cutoff scale, $\Lambda \sim M_{Pl} \,\g$.
This conjecture was used \cite{WGC} to rule out Extra Natural Inflation \cite{ExtraN}, where the inflaton arises from a Wilson line in a five-dimensional $U(1)$ gauge theory compactified on a circle.    

The Weak Gravity Conjecture might also be generalized to $D$ dimensions, $p$-form Abelian gauge fields, and their $p$ spacetime dimensional charged objects.  
Then, in a four dimensional gravitational theory with a 0-form axion, there must exist an instanton with action, $S_{cl} \lesssim M_{Pl}/f$,  where $f$ is the axion decay constant.  Although this conjecture lacks convincing motivation from black hole physics, the same phenomenon was observed in \cite{BDFG} in several diverse string theoretic setups.  If valid, it would essentially rule out single field models of inflation with super-Planckian axion decay constants, as instanton corrections would always introduce higher harmonics to the inflation potential:
\be
e^{n \frac{M_{Pl}}{f}} e^{i n \theta} \,,
\ee
effectively limiting the axion field range.  Analogous arguments considering gravitational instantons in effective field theory lead to similar conclusions \cite{UrangaGI}.  Moreover, the examples studied in \cite{BDFG} demonstrated the difficulty in obtaining axions with large decay constants within the limits of perturbative string theory, and raised the question if this is possible at all. 

Recent work has focused on whether these constraints from the Generalized Weak Gravity Conjecture on axion inflation can be evaded, in particular, by introducing multiple axion fields \cite{CR, sundrum, WGC1, WGC2, WGC3, WGC4, WGC5, WGC6, Eran}.  There has also been a large amount of work towards developing string theoretic models of large field inflation with sub-Planckian axion decay constants.  These come under two main classes,  firstly, stringy chaotic inflation scenarios where monodromy effects explicitly break the axion shift symmetry \cite{AM1, AM2, KS, KLS, MSU, Arthur, AM3, sebastian, HMRW}, and secondly, many field models where multiple axions generate an effective decay constant that is super-Planckian \cite{KNP, Nflation, CKY, LLM, SSY1, SSY2}.  Most constructions have used closed string axions in type II string theory.

In this letter, we revisit open string, single field models of axion inflation, in the light of quantum gravity constraints and the Weak Gravity Conjecture discussed above.  Open string inflatons include Wilson lines on wrapped D$p$-branes and the position moduli of moving D$p$-branes.  In fact, these scenarios are T-dual to each other.  By considering D3-branes moving down a long warped throat, Baumann and McAllister pointed out a rigid, sub-Planckian upper bound on the field range \cite{BM, BMrev}, and this can indeed be interpreted as a consequence of the Weak Gravity Conjecture.  However, single field models have been proposed with moderately super-Planckian decay constants.  Wilson lines on wrapped D$p$-branes with sub-Planckian decay constants were studied in \cite{WL}, and with super-Planckian decay constants in \cite{AZ}.  Planckian decay
 constants from wrapped D$p$-branes moving down or around a warped throat were found, respectively, in \cite{Becker} and \cite{KT}.  

We show that warping indeed allows for single field models with super-Planckian axion decay constant consistently with the Weak Gravity Conjecture.  The large decay constants are generated within the perturbative limits of the supergravity approximation.  Moreover, scalar potentials that break the continuous axion shift symmetry to a discrete one are potentially generated by classical or loop effects  \cite{ExtraN,KT,Hosotani,Antoniadis1,Antoniadis2}.  Therefore, any non-perturbative instanton effects are by construction exponentially suppressed, and would not rule out single field, large field, slow-roll inflation.

Unfortunately, these results do not lead to promising models of large field inflation and observable primordial gravitational waves.  This is because in explicit constructions, there is a tension between obtaining large decay constant and high string scale.  In fact, as we emphasize, it is always difficult to obtain a sufficiently high string scale within the limits of perturbation theory.  This presents an important challenge in building string theoretic models of large field inflation.

The paper is organized as follows.  In the next section we introduce inflation from D-branes in warped geometries and fix our conventions.  In Section III we study scenarios in which the candidate axionic inflaton is a Wilson line on a wrapped D-brane, and in Section IV we turn to the T-dual picture of the position modulus of a wrapped D-brane.  Finally, in Section V we discuss our results,  and the light they shed on the Generalized Weak Gravity Conjecture.

\section{Open String Inflation}

Our starting point is a generic type IIB string warped compactification from ten to four dimensions with metric %IZ: adding this:
(in the Einstein frame\footnote{Our conventions for going from string to Einstein frame are $G_{MN}^E = e^{\frac{\varphi_0-\varphi}{2}}G_{MN}^s$, where $\varphi$ is the dilaton, whose vev $\langle \varphi\rangle= \varphi_0$ defines the string coupling as $g_s = e^{\varphi_0}$. In these conventions the volumes evaluated in the background are frame independent. }):
\be
ds^2 = h^{-1/2}(r) g_{\mu\nu} dx^\mu dx^\nu  + h^{1/2}(r) g_{mn} dx^m dx^n \,,\label{E:10dmetric}
\ee
where $\mu, \nu = 0, \dots 3$, $m,n = 4, \dots 9$ and $h(r)$ is the warp factor, 
 possibly trivial, depending on a radial-like direction in the internal space. For example, for an $adS_5 \times X_5$ geometry which describes well a generic warped throat generated by branes and fluxes in the mid-throat region, we have:
\be
h(r) = L^4/r^4 \quad \textrm{and} \quad ds_6^2 = g_{mn} dx^m dx^n \!\!= \! dr^2 + r^2 d\Omega_5^2 \,,\label{E:adsthroat}
\ee

\noindent where $L$ is the adS length scale and $d\Omega_5^2=\tilde{g}_{ij} d\phi^i d\phi^j$ is the metric on some five-dimensional Einstein-Sasaki space that ensures ${\cal N}=1$ supersymmetry.  To construct a smooth, compact internal space out of the adS throat, we take $r_{IR} < r < r_{UV}$, where the adS region is glued to the tip of the throat at the IR cutoff and to a compact Calabi-Yau at the UV cutoff \cite{GKP}.

The four-dimensional Planck mass after compactification takes the form:
\be
M_{Pl}^2 =  \frac{4\pi \V_6^w}{g_s^2}  M_s^2  \quad \textrm{with} \quad \V_6^w l_s^6 = \int d^6y \sqrt{\det g_{mn}}\, h \,,
\ee

\noindent where we defined the string scale as $l_s^2=M_s^{-2} = (2\pi)^2\alpha'$, 
$g_s=e^{\varphi_0}$ is the string coupling and  $\V_6^w$ is the warped volume of the six-dimensional internal space in string units.  For example, assuming most of the volume comes from the middle region of an adS throat generated by $N$ D3-branes at its tip, we have:
\be
L^4 = \frac{g_s N}{4 \V_5}l_s^4  \label{E:L}
\ee
and
\be
\V_6^w l_s^6 = \frac12 \V_5  L^4 r_{UV}^2 \,,
\label{E:V6w}
\ee
with $\V_5=\int d\Omega_5$ the dimensionless volume of the base of the cone. 

To the above space we add a probe space-filling D$p$-brane wrapping a $(p-3)$-cycle in the internal space, with worldvolume coordinates $\xi^A$ ($A,B = 0, \dots, p+1$) and DBI action (in the Einstein frame):
\be
S_{DBI} = - T_p \int d^{p+1}\xi \,\sqrt{-\det\left(\gamma_{AB} + \mathcal{F}_{AB}\right)} \,.
\label{E:DBI}
\ee
The tension of the brane in the Einstein frame is given by:
\be
T_p = \mu_p\, g_s^{-1} \quad \textrm{with} \quad \mu_p = (2\pi)^{-p}(\alpha')^{-\frac{(p+1)}{2}}\,.
\ee
Also, $\gamma_{AB} = g_{MN} \partial_A X^M \partial_B X^N$ is the pullback of the ten-dimensional metric onto the brane ($M,N = 0, \dots 9$).  Finally, $\mathcal{F}_{AB} = \mathcal{B}_{AB} + 2\pi \alpha' F_{AB}$, with $\mathcal{B}_{AB}$ the pullback of the NSNS 2-form onto the brane, and $F_{AB}$ the field strength associated to the worldvolume gauge field.  We will choose static coordinates for the brane, so $\xi^A= (x^\mu, y^a)$ with $y^a$ the $(p-3)$ internal coordinates along the brane. 

Possible open string inflaton fields in the above system include Wilson line moduli associated with the worldvolume gauge field \cite{WL,AZ} and the moduli describing the position of the D$p$-brane in the compact space \cite{DT,BMNQRZ,KKLMMT}. 
We will consider single field, large field inflation and the Weak Gravity Conjecture in these setups, which are related to each other by T-duality.

\section{Wilson Line Inflation}

When the $(p-3)$-cycle wrapped by the D$p$-brane contains a non-trivial $1$-cycle\footnote{Note that this does not require the full six-dimensional internal space to have a 1-cycle, see \cite{KKLAM} for examples.}, parameterized by some coordinate $\phi$,   the brane can have a Wilson line wrapping its worldvolume:
\be\label{WL}
e^{i \theta} = e^{i\oint A_\phi d\phi}  \,.
\ee

Upon dimensional reduction, the DBI action (\ref{E:DBI}) for the brane in the background (\ref{E:10dmetric}-\ref{E:adsthroat}) includes a 4D gauge kinetic term for the $U(1)$ worldvolume gauge field:
\be
S=-\int d^4x \sqrt{-g} \, \frac{1}{4\g_4^2} \,F_{\mu\nu} F^{\mu\nu} \,,\label{E:U1action}
\ee
where the four-dimensional effective gauge coupling constant, $\g_4$, is computed to be:
\be
\g_4^2 = (2\pi) g_s h_0^{(3-p)/4} \,(n \,\V_{p-3})^{-1} \,. \label{E:U1g}
\ee 
Here, $h_0$ 
is the value of the warp factor evaluated at the brane position  
and $\V_{p-3}$ is the unwarped volume of the $(p-3)$-cycle wrapped by the brane in units of $l_s$, $\V_{p-3} l_s^{p-3} = \int d^{p-3}y \sqrt{\det g_{ab}}$ (notice that $\V_{p-3}$ also depends on the position of the brane, see eq.~\eqref{E:adsthroat}). Finally $n$ is the wrapping number. 

At the same time, the DBI action leads to a kinetic term for the Wilson line modulus, $\theta=2\pi A_\phi$, which takes the form  
\be
S = -\int d^4x \sqrt{-g} \,\, \frac{f^2}{2} \partial_\mu \theta \partial^\mu \theta \label{E:Saxion}
\ee
where the axion decay constant, $f$, is computed to be: 
\be
f^2 = \frac{n \,g_s^{-1}}{(2\pi)^3} \frac{h_0^{(p-7)/4} \V_{p-3}}{R_0^2}
\ee 
Here, $2\pi R_0$ is the unwarped length of the 1-cycle with line element $ds^2 = R_0^2 \,d\phi^2 = r_0^2 \tilde{g}_{\phi\phi} d\phi^2$.

The axion, $\theta$, has a continuous shift symmetry descending from the higher dimensional gauge symmetry, which may be broken to a discrete one. 
 A slow-roll potential can thus be generated for the Wilson line axion, $\theta$, by fluxes, brane backreaction, warping, loop contributions and/or other effects, including non-perturbative ones \cite{ExtraN, BDKMM, WL,MRZ}.   
 Large field excursions for $\theta$ are encoded in the field's decay constant by $f/M_{Pl} >1$, with:
\be
\frac{f^2}{M_{Pl}^2} = \frac{ g_s }{2(2\pi)^{4}}  \frac{h_0^{(p-7)/4} l_s^2}{R_0^2} \frac{n\,\V_{p-3}}{\V_6^w} \,.\label{E:WL-Largef}
\ee

As a first example, consider a D$p$-brane in an unwarped, possibly anisotropic toroidal orientifold compactification, with $j$ directions of size $R$, and $6-j$ directions of size $L$.  Taking the brane to wrap a $(p-3)$-cycle with $j_b$ directions of size $R$ (so $j \geq j_b$) and $p-3-j_b$ directions of size $L$ (so $6-j \geq p-3-j_b$), and the Wilson line wrapping one of the $R$-directions (so $j_b \geq 1$), we have:
\be
\frac{f^2}{M_{Pl}^2} = \frac{ n \,g_s }{2(2\pi)^{4}(2\pi)^{9-p}}  \left(\frac{l_s}{R}\right)^{2+j-j_b} \left(\frac{l_s}{L}\right)^{9-p-j+j_b}  \,.\label{E:WLtorus-Largef}
\ee 
A large $f/M_{Pl}$ would require a substring scale cycle\footnote{Vanishing cycles do occur in more general string geometries, for example when blowing-up a singularity.  The vanishing cycles in the conifold were used in \cite{Witten06} to achieve small axion decay constants from string theory, and in \cite{Grimm} to achieve string axion $N$-flation models with near Planck scale axion decay constants.} $R<l_s$ and/or $L<l_s$.  In this limit the perturbative description breaks down, and the T-dual description should be used. 

Next, consider a wrapped D5-brane in an $adS_5 \times X_5$ warped throat.  The axion decay constant is then:
\be
\frac{f^2}{M_{Pl}^2} = \frac{g_s}{(2\pi)^4} \frac{n \V_2}{\V_5}  \frac{l_s^6}{L^6} \frac{r_0^2}{r_{UV}^2} \frac{l_s^2}{R_0^2}\,.\label{E:WLD5-Largef}
\ee
Since $\V_2$ goes as $r_0^2/l_s^2$, the possibility now arises that $f/M_{Pl} >1$ can be achieved with a brane at the top of a long throat, $r_0 \sim r_{UV}$ and $r_{UV}\gg L$.  

Now let us check whether the conditions for large field inflation can be fulfilled consistently with the Weak Gravity Conjecture.  Indeed, probe D-branes can be used to test the consistency of a string theory configuration by asking whether  their worldvolume gauge theories and charged matter satisfy the Weak Gravity Conjecture, wherever they are placed. The Weak Gravity Conjecture implies a new UV cutoff, $\Lambda = M_{Pl}\, \g_4$, which in the present case is:
\be
\Lambda^2 = \frac{(2\pi)^{2}}{g_s\,l_s^2} \frac{2}{h_0^{(p-3)/4}} \frac{\V_6^w}{n\,\V_{p-3}} \,.\label{E:Lambda}
\ee
Matter fields charged under the $U(1)$ arise from strings stretching between the Wilson line D$p$-brane and a separated parallel D$p$-brane.  Largest masses arise when the branes are very distant, with $m \sim M_s^2 \ell_\Phi$ and $\ell_\Phi$ the brane separation. 
So the Weak Gravity Conjecture, $m^2 < M_{Pl}^2\, \g_4^2$ requires:
\be
1 < \frac{(2\pi)^{2}}{g_s} \frac{l_s^2}{\ell_{\Phi}^2} \frac{2}{h_0^{(p-3)/4}} \frac{\V_6^w}{n\,\V_{p-3}}\,. \label{E:WL-WGC}
\ee
The Weak Gravity Conjecture  (\ref{E:WL-WGC}) then imposes an upper limit on $f/M_{Pl}$  (see eq. (\ref{E:WL-Largef})):
\be
\frac{f^2}{M_{Pl}^2} < \frac{h_0^{-1}}{(2\pi)^2}\frac{l_s^4}{\ell_\Phi^2 R_0^2} \,.
\ee

Considering an $adS_5 \times X_5$ throat \eqref{E:adsthroat} as a prototype, 
the warped length of a string stretching from a D$5$-brane at $r_0$ near the top of the throat all the way down to the bottom of the throat is:
\be
\ell_\Phi \sim \int^{r_0}_{r_{IR}} h^{1/4} dr = L \ln \left(\frac{r_0}{r_{IR}}\right)\,. \label{E:lPhi}
\ee
Taking $r_{IR} \sim l_s$ and $r_0 \sim r_{UV}$ we then obtain from  eq. (\ref{E:WL-WGC}):
\be
1 < \frac{(2\pi)^2}{n g_s}\frac{r_0^2/l_s^2}{(\ln(r_0/l_s))^2} \frac{r_0^2 \,\V_5}{l_s^2\,\V_2} \,,
\ee
where again recall that $\V_2$ goes as $r_0^2/l_s^2$.
Note in particular that the Weak Gravity Conjecture does not put an upper limit on $r_0$, so both the Weak Gravity Conjecture and large axion decay constant could indeed be satisfied with a brane at the top of a long throat.

It should be also  checked that the backreaction of the wrapped probe D$5$-brane can be safely neglected. 
This can be estimated as follows \cite{Becker}. 
 The local contribution from a D$p$-brane to the traced Einstein's equation goes as \cite{GKP,Becker}:
\be
\left( T^m_m - T^\mu_\mu \right)^{loc} = (7-p) T_p \Delta^{(9-p)}(\Sigma_{p-3})
\ee
where $\Delta^{(9-p)}(\Sigma_{p-3})=\delta^{(9-p)}(\Sigma_{p-3})/\sqrt{\det g_{9-p}}$ is the covariant delta function on the wrapped $(p-3)$-cycle, $\Sigma_{p-3}$, with transverse volume $\int \sqrt{\det g_{9-p}}$.  The condition that the backreaction of the wrapped D5-brane is negligible can then be written as:
\be
\frac{n}{2N} \frac{T_5}{T_3}\frac{\Delta^{(4)}(\Sigma_{2}) }{\Delta^{(6)}(\Sigma_{0})} \ll 1 \,. \label{E:bkreaction}
\ee 

For example, for the D5-brane in the $adS_5\times X_5$  background, we can take the parameters $L\sim \sqrt{3}\, l_s$, $g_s \sim 0.3$, $r_0 \sim r_{UV} \sim 500 \,l_s$, $R_0 = r_0 \,\tilde{g}_{\phi\phi}^{1/2} \sim l_s$, $n \sim 100$, $\V_2l_s^2 \sim \pi r_0^2$ and $\V_5 \sim \pi^3$. 
To compute the backreaction, we estimate transverse volume elements as $\sqrt{\det g_4} \sim h_0 r_0^3 \sin\theta_1$ and $\sqrt{\det g_6} \sim h_0^{3/2} r_0^5 \sin\theta_1\sin\theta_2$, for some internal coordinate angles $\theta_1, \theta_2$.  Thus, we can consistently achieve $ f \sim 4.2 \,M_{Pl}$, $M_s \sim 1.4 \times 10^{-5} M_{Pl}$ and $\Lambda \sim 4.5 \times 10^{-2} M_{Pl}$.  It is also conceivable that more complicated warped geometries than the adS mid-throat geometry may allow one to achieve higher $f/M_{Pl}$ safely within the limits of perturbation theory.

To summarise, it is plausible that  a Wilson line on a wrapped D$5$-brane can give rise to an axion with large axion decay constant consistently with the Weak Gravity Conjecture. 
This can be achieved by ensuring a warped throat geometry with a 1-cycle inside the $2$-cycle that is wrapped by the D$5$-brane in a region with small warp factor, $h_0^{1/2} \ll 1$.  
A concrete realisation of the proposal would require a throat geometry with these properties.  
It would likely not, however, correspond to a stringy realisation of the Extra-Natural Inflation proposal \cite{ExtraN}, as the string scale  would probably turn out to be too low. 
Indeed, note that taking $r_0 \sim r_{UV}$ large to achieve large $f/M_{Pl}$ drives the string scale to smaller values.
 We will comment further on this below.  First we turn to the T-dual description of a D-brane Wilson line, which is a D-brane position field.  In this case, more explicit constructions are possible.

\section{Brane Position Inflation}

Once again consider a space-filling D$p$-brane wrapping a $(p-3)$-cycle, and now assume that the position of the brane in one of the compact angular dimensions corresponds to an inflaton field.  Such scenarios have been studied, for example, in \cite{GBRZ,GRZ, Becker, KT,Arthur}.  Dimensional reduction of the DBI action (\ref{E:DBI}) gives, in exactly the same way as for the Wilson line scenario, a $U(1)$ gauge group in the four-dimensional effective field theory, see (\ref{E:U1action}), (\ref{E:U1g}).  It also leads to a kinetic term for the angular position modulus, $\theta$, of the form (\ref{E:Saxion}).  %, 
The axion decay constant, $f$, is now given by:
\be
f^2 = \frac{(2\pi) \,g_s^{-1}}{l_s^4} h_0^{(p-3)/4} g_{\theta\theta} \,n\,\V_{p-3} \,. \label{E:PO-f}
\ee
A large axion decay constant requires $f/M_{Pl} >1$ with:
\be
\frac{f^2}{M_{Pl}^2} = \frac{1}{2} \,g_s h_0^{(p-3)/4} g_{\theta\theta} \frac{n\,\V_{p-3}}{l_s^2\V_6^w} \,.\label{E:PO-Largef}
\ee
The axion, $\theta$, corresponds to a periodic direction and so inherits a discrete shift symmetry.  Again, the inflaton potential may be induced by fluxes, brane backreaction, warping, loop contributions and/or other effects, including subleading non-perturbative effects.  For instance, \cite{KT} found a potential of the Natural Inflation form (\ref{E:NI-V}), by considering the forces experienced by a D-brane moving in the geometry of the warped resolved conifold glued to a compact Calabi-Yau.  

The new UV cutoff implied by the Weak Gravity Conjecture is the same as for the Wilson line scenario, see (\ref{E:Lambda}).  Charged matter may again arise due to strings stretching to 
 parallel D-branes that are separated by a distance $\ell_\Phi$ along the throat or around the $\theta$-direction.  
The Weak Gravity Conjecture, $m < \g_4 M_{Pl} $ constraint is the same as before, eq.~\eqref{E:WL-WGC}, and implies the following upper bound on the axion decay constant,%
\footnote{We thank Zac Kenton and Steve Thomas for pointing out an error in the previous version of this equation.} \eqref{E:PO-Largef}:
\be\label{E:PO-WGCfinal}
\frac{f^2}{M_{Pl}^2}< (2\pi)^2  \, \frac{g_{\theta\theta}}{\ell_\Phi^2} \,.
\ee

We can now consider these constraints in some explicit constructions.  Take first a possibly anisotropic unwarped toroidal orientifold compactification, with $j$ directions of large size $R$ and $6-j$ directions of small size $L$.  Consider a D$p$-brane wrapping $j_b$ large directions and $p-3-j_b$ small directions, cycling a large direction (so $j_b \leq j-1$).  The position modulus has axion decay constant given by:
\be 
\frac{f^2}{M_{Pl}^2} = \frac{n\,g_s }{2} \frac{1}{(2\pi)^{9-p}}  \left(\frac{R}{L}\right)^{2+j_b-j} \left(\frac{l_s}{L}\right)^{7-p} \,.\label{E:PO-torus-f}
\ee
So a large decay constant could be achieved with a large angular direction $R\gg L$ for $j-j_b=1$.  We can test the consistency of the required parameters by using the Weak Gravity Conjecture.  To this end we introduce another probe D$p$-brane, separated by a distance $\ell_\Phi \sim \pi R$ around a large angular direction.  The Weak Gravity Conjecture (\ref{E:WL-WGC}) then requires:
\be
1 < \frac{8}{n \,g_s} (2\pi)^{9-p} \left(\frac{L}{R}\right)^{2+j_b-j} \left(\frac{L}{l_s}\right)^{7-p}
\ee
and so imposes an upper bound on the decay constant:
\be
\frac{f^2}{M_{Pl}^2} < 4  \,, \label{E:torus-fWGC}
\ee
which is at most $f \sim 2 \, M_{Pl}$.  For example, taking a D3-brane, with  $g_s\sim 0.3$, $L \sim l_s$, $R \sim 1.5 \times 10^6 \, l_s$, $j=1$, $j_b=0$ and $n=1$, we obtain $f \sim 1.9 \, M_{Pl}$, $M_s \sim 2.8 \times 10^{-7} M_{Pl}$ and $\Lambda = 1.4 \times M_{Pl}$.  Note that to achieve $f \sim 2 \, M_{Pl}$, the  string scale is very low.

Let us next consider a $Dp$-brane moving in a warped throat.  The prototypical example is the Klebanov-Strassler warped throat, produced by placing $N$ D3-branes at the tip of the deformed conifold. 
Away from the conical deformation, in the mid-throat region, the base of the conifold can be taken to be $T^{1,1}$, and the 10D metric takes the form \eqref{E:10dmetric}, with $h=L^4/r^4$ and
\bea
ds_6^2 &=& dr^2 +\frac16 r^2 \left(d\theta_1^2 + \sin^2\theta_1 d\phi_1^2\right) \nonumber \\
&&  + \frac16 r^2 \left(d\theta_2^2 + \sin^2\theta_2 \d\phi_2^2\right) \nonumber \\
&&+ \frac19 r^2 \left(d\psi^2 + \cos\theta_1 d\phi_1 + \cos\theta_2 d\phi_2 \right)^2 \,.
\eea
Allowing for a $\Z_k$ orbifolding in the $\psi$ direction, the volume of $T^{1,1}$ is given by 
${\cal V}_5=\frac{16}{27k}\pi^3$ and the $adS$ radius of curvature (\ref{E:L}) is:
\be
L^4 = \frac{27 k}{64 \pi^3} g_s N l_s^4 \,.\label{E:RC-L}
\ee
Now consider a D5-brane wrapping a 2-cycle in the throat
\footnote{Note that for $p=3$, the warp factor cancels out in $f/M_{Pl}$, eq. (\ref{E:PO-Largef}), and cannot be used to make $f/M_{Pl}$ large.} \cite{Becker}:
\be
r=r_0, \; \psi = \psi_0, \; \theta_1 = -\theta_2, \; \phi_1 = -\phi_2 \label{E:2-cycle}
\ee
and moving in an angular direction, say, $\theta_2$.

The condition that the backreaction of the probe D5-brane can be consistently neglected, eq. (\ref{E:bkreaction}), evaluates in the present case to\footnote{The volume orthogonal to the D5-brane embedded into spacetime with the static gauge is obtained by setting $\theta_1=\phi_1=0$.}:
\be
\frac{n}{12 N}\frac{L^2}{l_s^2} \sin\theta_1 \ll 1 \,.
\ee

The parameters are also constrained by the Weak Gravity Conjecture.  To see how, we introduce another probe D$p$-brane, firstly separated from $r_0$ up the throat to $r_{UV}$, and secondly separated around the throat along $\theta_2$ at $r=r_0$.  We estimate the mass of the corresponding charged matter, $m = M_s^2 \ell_\Phi$, as (c.f. (\ref{E:lPhi})):
\bea
&&m = M_s^2 L \log \frac{r_{UV}}{r_0} \,, \label{KS-mass-r}\\
&&m = M_s^2 \frac{\pi L}{\sqrt{6}} \,,\label{KS-mass-theta}
\eea
where we have taken into account that the lengths of the stretched strings, $\ell_\Phi$, are warped.  The Weak Gravity Conjecture (\ref{E:WL-WGC}) thus implies that (use eqs. (\ref{E:V6w}), (\ref{E:2-cycle}), $\V_2 l_s^2 = \frac{4\pi}{3}r_0^2$ and $h_0=L^4/r_0^4$):
\be
1 < \frac{32 \pi^2}{3}\frac{1}{n k \,g_s}\frac{r_{UV}^2}{l_s^2} \,, \label{E:SC-WGC} 
\ee
which gives the following upper bound on $f$:
\be
\frac{f^2}{M_{Pl}^2} < \frac{4  r_0^2}{L^2}\,.
\ee
Comparing this with result for the torus (\ref{E:torus-fWGC}), we see that warping has relaxed the upper bound on the axion decay constant, and  larger super-Planckian values may now be possible within the supergravity approximation.

In fact, one can immediately infer from Eqs (\ref{E:PO-Largef}), (\ref{E:RC-L}), (\ref{E:V6w}), $g_{\theta\theta}=r_0^2/6$, $\V_2 l_s^2 = \frac{4\pi}{3}r_0^2$ and $h_0 = L^4/r_0^4$, that the axion decay constant is given by:
\be
\frac{f^2}{M_{Pl}^2} = \frac{3 \,  n k\, g_s}{8\pi^2} \frac{l_s^2 r_0^2}{L^2 r_{UV}^2}
\ee 
As $r_0 \lesssim r_{UV}$ and $L > l_s$, in the end only the orbifold and wrapping numbers can help to increase $f$.  However, moderate super-Planckian decay constants are attainable at the limits of the supergravity approximation and small brane backreaction\footnote{Larger values for $f/M_{Pl}$ are possible compared to \cite{Becker} by considering the brane moving in an angular direction rather than the radial direction.}, consistently with the Weak Gravity Conjecture.  
For example, taking $g_s\sim 0.3$, $L \sim \sqrt{3} \, l_s$, $r_{UV}\sim r_0 \sim 30 \, l_s$, $n \sim 30$ and $k\sim 100$, we obtain $f \sim 3.4 \, M_{Pl}$, $M_s \sim 3.1 \times 10^{-3} M_{Pl}$ and $\Lambda = 7.1 \times 10^{-2} M_{Pl}$.  
Again, the string scale results to be quite low (c.f. eq.~\eqref{E:Vinf}).

More complicated warped geometries may allow for even larger axion decay constants.  For example, a D5-brane moving in the warped resolved conifold was used in \cite{KT} to obtain a super-Planckian decay constant, and proposed as a model for Natural Inflation.   The 10D metric for the warped resolved conifold is \cite{PZT,KM}:
\be
ds_{10}^2= h(r,\theta_2)^{-1/2} ds_4^2 
 + h(r,\theta_2)^{1/2} ds_6^2\,,
\ee
with
\bea
ds_6^2 &=& \kappa^{-1}(r) dr^2 +\frac16 r^2 \left(d\theta_1^2 + \sin^2\theta_1 d\phi_1^2\right) \nonumber\\
&&  + \frac16 \left(r^2 + 6u^2\right) \left(d\theta_2^2 + \sin^2\theta_2 \d\phi_2^2\right) \nonumber\\
&&+ \frac19 \kappa(r) r^2 \left(d\psi^2 + \cos\theta_1 d\phi_1 + \cos\theta_2 d\phi_2 \right)^2 \!,
\eea
where
\be
\kappa(r) = \frac{r^2 + 9u^2}{r^2 + 6 u^2}\,
\ee
and $u$ is the resolution parameter. The warp factor $h(r,\theta_2)$ is given by:
\be
h(r,\theta_2) = L^4 \sum_{l=0}^{\infty} (2l+1) H_l(r) P_l(\cos\theta_2) \,,
\ee
where
\be
H_l(r)=\frac{2}{9 u^2} \frac{C_\beta}{r^{2+2\beta}} \; {}_2F_1\left(\beta,1+\beta,1+2\beta;-\frac{9u^2}{r^2}\right)
\ee
and
\be
C_\beta = \frac{(3u)^{2\beta}\Gamma(1+\beta)^2}{\Gamma(1+2\beta)} \quad \textrm{and} \quad \beta=\sqrt{1+\frac32 l (l+1)} \,.
\ee
The warped geometry is sourced by $N$ D3-branes placed at the tip of the throat, at $r=0$ and $\theta_2=0$.  Very close to the branes, at $\theta_2=0$ and small $r$, the geometry corresponds to $adS_5 \times S^5$ with $h=L^4/r^4$ and $L$ given in \eqref{E:RC-L}.  Further away from the branes, but still at small $r < u$, the warp factor is $h= L^4/(r u)^2$.  Taking the smeared limit for the $N$ source D3-branes, the warp factor can be approximated by the $l=0$ mode as \cite{PZT,KM}: 
\be
h(r) = \frac{2 L^4}{9 r^2u^2} -\frac{2 L^4}{81 u^4} \log\left(1 + \frac{9 u^2}{r^2} \right) \,.
\ee
Using this approximation to compute the warped volume, one obtains
\be
V_6^w = s \frac{64 \pi^3}{81 k} L^4 r_{UV}^2
\ee
with
\bea
s=&&2  + \frac{1}{9} \frac{r_{UV}^2}{u^2} \left(1-\log\left(1+\frac{9u^2}{r_{UV}^2}\right) \right) \nonumber \\
&&  - \frac{1}{81} \frac{r_{UV}^4}{u^4}\log\left(1+\frac{9u^2}{r_{UV}^2}\right) \,.
\eea
The string scale is then given by:
\be \label{E:RC-Ms} 
M_s^2 = M_{Pl}^2 \frac{81 k}{4(2\pi)^4 s}\frac{g_s^2\, l_s^6}{L^4 r_{UV}^2} \,.
\ee
In the near-tip region where the warp factor is $h_0= L^4/(r_0 u)^2$, $g_{\theta\theta} = u^2$ and the wrapped volume is $\V_2 l_s^2 = 4\pi u^2$, taking $u \sim r_{UV}$, the axion decay constant becomes:
\footnote{This result differs from that in \cite{KT}  since we use $h = \frac{L^4}{(r u)^2}$ valid for $\theta_2 \neq 0$, in contrast to 
$h = \frac{L^4}{r^4}$, used in \cite{KT}. This latter warp factor is also used to compute the warped volume of the throat up to the UV cutoff in \cite{KT}.}
\be
\frac{f^2}{M_{Pl}^2} = \frac{6561\, n k\,g_s}{32\pi^2(171-10\log 10)}\frac{l_s^2 u}{L^2 r_0} \,.\label{E:RC-Largef}
\ee
The possibility now emerges that small $r_0 < u$ could correspond to super-Planckian values \cite{KT}.

The condition that the backreaction of the probe D5-brane be much smaller than that of the $N$ D3-branes \eqref{E:bkreaction} can now be written as:
\be
\frac12 \frac{n}{N} \frac{L^2}{l_s^2} \frac{r}{u} \sin\theta_1 \ll 1 \,. \label{E:RC-bkreaction}
\ee
Also, the Weak Gravity Conjecture (\ref{E:WL-WGC}) now implies that:
\be
1 < \frac{128(171-10\log10)\pi^2}{6561} \frac{1}{g_s n k}\frac{r_0^2}{l_s^2}. \label{E:RC-WGC} 
\ee
where we used that the masses of warped strings stretching between probe D-branes separated up the throat and around the throat are, respectively (c.f. \eqref{KS-mass-r}, \eqref{KS-mass-theta}):
\bea
&&m = M_s^2 \int_{r_0}^{r_{UV}} d r h^{1/4} \kappa^{-1/2} \,, \\
&&m = M_s^2 \pi L \frac{u^{1/2}}{r_0^{1/2}}\,.
\eea
Notice, for example, that the Weak Gravity Conjecture (\ref{E:RC-WGC}) ensures that the wrapping number, orbifold number and string couplings are not too large, as expected.  It leads to a simple upper bound on the axion decay constant (see eq. \eqref{E:RC-Largef}):  
\be
\frac{f^2}{M_{Pl}^2}< \frac{4 \,r_0 u}{L^2} \,. \label{E:RC-mf}
\ee 

It is now possible to achieve large $f/M_{Pl}$ within the limits of perturbative control, small brane backreaction and consistently with the Weak Gravity Conjecture.  For example, taking $g_s \sim 0.3$, $L \sim \sqrt{3} \, l_s$, $u \sim r_{UV} \sim 65 \, l_s$, $r_0 \sim 2 l_s$, $n \sim 10$ and $k\sim 10$, we find $f \sim 6.8 \, M_{Pl}$, $M_s \sim 2.1 \times 10^{-4} M_{Pl}$ and $\Lambda = 1.2 \times 10^{-2} M_{Pl}$, consistently with \eqref{E:RC-bkreaction} and \eqref{E:RC-WGC}.

Although we have seen that warping allows one to consistently obtain axions with large axion decay constant in string theory, these axions cannot be used for large field inflation because the corresponding string scales are too small.  Large field inflation for observable tensor modes implies that inflation occurs at least at GUT scale $M_{inf} \sim 10^{-2} M_{Pl}$ (see eq. (\ref{E:Vinf})), so the string scale should be at least around $M_s \sim 10^{-1} M_{Pl}$, for a supergravity analysis to be valid during inflation\footnote{Strictly speaking, a hierarchy of scales $M_{inf} < M_c <  M_s \lesssim M_{Pl}$, where $M_c$ is the compactification scale,  is needed in order to consistently work in the four dimensional, low energy supergravity limit of string theory \cite{BMrev}, and moduli must also be stabilised.}.  In fact, as is well known, dimensional reduction gives the following relation between the string scale and Planck scale in any string model:
\be
M_s^2 = M_{Pl}^2 \frac{g_s^2}{4\pi \V_6^w} \,.
\ee
Thus a string scale $M_s \gsim 10^{-1} M_{Pl}$ is difficult to achieve within the limits of supergravity.  This renders questionable all models of large field inflation in perturbative string theory, even those with sub-Planckian decay constants like axion monodromy\footnote{Factors like $(2\pi)$ might help in general to achieve higher string scales within the perturbative limits,  but  explicit constructions enhance this problem.  For example, in the original axion monodromy proposal \cite{AM3}, increasing $\V_6^w$ by using throats within throats to prevent brane anti-brane annihilation and suppress brane backreaction \cite{bifid1, bifid2} will drive the string scale down.  The Large Volume Scenario used in the D7-brane chaotic inflation model of \cite{Arthur} would also make a high string scale difficult to achieve. See also \cite{Blumenhagen}.}.

\section{Discussion}

We have shown that single open string axions coming from D-branes in warped geometries can enjoy super-Planckian axion decay constants consistently with the Weak Gravity Conjecture and within the limits of the supergravity approximation. 
The most explicit realisation is the model proposed by Kenton and Thomas \cite{KT} for Natural Inflation, using the position modulus of a D-brane moving in a compact angular direction of the warped resolved conifold.  However, none of the constructions considered here can be used for large field inflation, as their string scales are too low compared to the inflationary scale \eqref{E:Vinf}. 
 Indeed, we stress that one of the biggest challenges in realising large field inflationary models within string theory is to identify appropriate compactifications with sufficiently high string scale, within the limits 
of four dimensional supergravity. One might speculate that any future observation of primordial gravitational waves could be an indication that the strongly coupled regime of string theory is most appropriate for describing Early Universe Cosmology.

Although the open string axions with large decay constant studied here cannot be used for large field inflation, they may give some insight into the Generalized Weak Gravity Conjecture.  The D-brane axions have interpretation as (or are related to) gauge fields in higher dimensions, and so would be directly subject to the Weak Gravity Conjecture\footnote{See the contemporaneous paper \cite{HRR} for a further discussion on the Weak Gravity Conjecture and dimensional reduction and the Generalized Weak Gravity Conjecture for axions.}.  Furthermore, the Generalized Weak Gravity Conjecture, in its mild form, states that among any instantons, $i$, coupling to an axion with coupling $f_i$, there must exist at least one with action $S^i_{cl} \lesssim M_{Pl}/f_i$.  The strong form states, additionally, that this instanton must be the one with smallest action.  

However, we have seen that for open strings in warped geometries a large axion decay constant is generated within the limits of supergravity, where instanton effects must be exponentially suppressed.   Indeed, the known instantons that couple to the D$p$-brane Wilson line and position open string moduli are Euclidean E$q$-branes wrapping $(q+1)$-cycles on the internal space.  These  instantons usually couple to the D-brane moduli with axion decay constant $f$, and their actions go as $g_s^{-1}$ and the volume of the wrapped $(q+1)$-cycle\footnote{See \cite{Pablo} for an explicit computation of the instanton generated scalar potential for open string moduli in toroidal orbifold compactifications with magnetized D-branes.}.  Thus they have $S_{cl} > M_{Pl}/f$, and contributions are exponentially suppressed within the supergravity approximation.  There then seems to be three likely possibilities: 
\textit{(i)} In the presence of warping, the E$q$-instantons couple to the open string axions with a suppressed effective axion coupling $f' < f$, such that they satisfy the Generalized Weak Gravity Conjecture  with large action.  \textit{(ii)} There are new stringy instanton effects that couple to the D-brane moduli with a suppressed effective axion coupling $f' < f$, which satisfy the Generalized Weak Gravity Conjecture\footnote{This is similar -- but slightly different -- to the loophole discussed in \cite{WGC2} to achieve Natural Inflation consistently with the strong form of the Generalized Weak Gravity Conjecture.} with large action, 
$S'_{cl}\lesssim M_{Pl}/f'$, either mildly ($S'_{cl} > S_{cl}$) or strongly ($S'_{cl} < S_{cl}$).   \textit{(iii)}   The Generalized Weak Gravity Conjecture is incorrect.  We hope this letter will help to shed light on these issues.

\section*{Acknowledgments}
We would like to  thank  Lilia Anguelova, Nima Arkani-Hamed, Nana G. Cabo B., Zac Kenton, Carlos N\'u\~nez, Fernando Quevedo, Gianmassimo Tasinato, Steve Thomas and Irene Valenzuela  for useful discussions.  
IZ thanks the Gordon Research Conference on String Theory and Cosmology  for partial support, hospitality and the stimulating atmosphere that inspired this project. She also thanks the Physics Department of the University of  Guanajuato, Mexico for hospitality while part of this work was done.  The research of SLP is supported by a Marie Curie Intra European Fellowship within the 7th European Community Framework Programme.

%\bibliography{refs}
%\bibliographystyle{utphys}

\providecommand{\href}[2]{#2}\begingroup\raggedright\endgroup

\end{document}